\documentclass[twocolumn,showpacs,preprintnumbers,amsmath,amssymb]{revtex4}
\usepackage{graphicx}% Include figure files
\usepackage{dcolumn}% Align table columns on decimal point
\usepackage{bm}% bold math

\usepackage{hyperref}

\newcommand{\beq}{\begin{equation}}
\newcommand{\eeq}{\end{equation}}
\newcommand{\bespl}{\begin{split}}
\newcommand{\enspl}{\end{split}}
\newcommand{\del}[1]{\frac{\partial}{\partial #1}}
\begin{document}

\preprint{}

\title{Bifurcation and stability of uniformly rotating homogeneous ellipsoids surrounded by a massive thin ring}% Force line breaks with \\

\author{Shin'ichirou Yoshida}
\email{yoshida@ea.c.u-tokyo.ac.jp}
\affiliation{
Department of Earth Science and Astronomy, Graduate School of Arts and Sciences,
The University of Tokyo\\
Komaba 3-8-1, Meguro-ku, Tokyo 153-8902, Japan
}%

\date{\today}% It is always \today, today,

\begin{abstract}
We examine the effects of a massive concentric ring around a spheroid or an ellipsoid with uniform density and uniform rotation. Equilibrium sequences of axisymmetric Maclaurin-like spheroid and triaxial Jacobi-like ellipsoids are obtained. Due to the gravitational field of the ring, Maclaurin-like spheroid does not have a spherical limit when the object's angular frequency vanishes. At a critical value of the eccentricity of the spheroid's meridional section, a triaxial Jacobi-like ellipsoid bifurcates. When a parameter characterizing the gravitational field of the ring is smaller than a threshold, the bifurcation points of Maclaurin-like and Jacobi-like ellipsoids exist and the critical eccentricity is slightly larger than that of the classical Maclaurin-to-Jacobi bifurcation. When the parameter exceeds the threshold, the Maclaurin-like spheroid does not have the bifurcation point and the Jacobi-like ellipsoid appears at the lower eccentricity than the Maclaurin-like spheroid. By comparisons of the energy of the ellipsoids with the same angular momentum, it is shown that the critical point of bifurcation does not correspond to the onset of the secular instability of Maclaurin-like spheroid. It is concluded that the gravitational field of a massive ring surrounding a uniformly rotating spheroid stabilizes it against a bar-shaped deformation due to viscous
dissipations.
\end{abstract}

\maketitle

%%%%%%%%%%%%%%%
\section{Introduction}
%%%%%%%%%%%%%%%
Rotating equilibrium figures of self-gravitating fluid have a long history of studies
since Isaac Newton's time. The studies have been motivated by the pursuits
of possible configurations
of planets, stars, stellar clusters, and galaxies, which span many orders of magnitude
in physical sizes and timescales. The simplest and still useful models are the
uniform-density and uniformly rotating spheroids and ellipsoids, which enable analytic expressions
of physical characteristics (see \citealt{2013srlm.book.....L,1969efe..book.....C} for the classical
results). 
%-- dwarf planet Haumea
These simple results are still useful in, for instance, studying the
equilibrium figure of a dwarf planet, Haumea, which is thought to have
a triaxial ellipsoidal shape \citep{2017Natur.550..219O,Dunham_2019,2022PSJ.....3..225N}.
%-- compact stars
In a completely different context of compact star physics, the models have been utilized
to mimic the rapidly rotating compact stars, which may assume time-dependent triaxial
ellipsoid. Then the objects emit gravitational waves whose frequencies are
optimal for the ground-based gravitational-wave observatories \citep{LVKprospects}.
 For some of the recent applications of the classical ellipsoids to compact stars, see, e.g., \cite{2022MNRAS.509.1854R,2022MNRAS.511.1942Y, 2022PhRvD.106h3004R}.
%-- stellar clusters, galaxies
At a larger scale, the classical ellipsoids are applied to model star clusters
\citep{2020ApJ...904..171I} or elliptical galaxies surrounded by dark matter 
haloes \citep{2015BaltA..24..408K}.
In fact, closely related studies to ours have been done by Kondratyev and his collaborators \citep{2010Ap.....53..189K,2015BaltA..24..408K,2016ARep...60..526K},
where the effects of a circum-galactic dark matter ring on the elliptical galaxies are investigated
to explain the origin of the highly flattened Early-type galaxies. 
An apparent lack of rotational velocity in some of the highly-flattened ellipsoidal 
galaxies are attributed to the tidal effect of a massive concentric ring around a 
uniformly rotating ellipsoid. 

%-- this paper
In this paper, we use the same model as that of Kondratyev and his collaborators
to study more fundamental issues of ellipsoidal figures. 
%-- bifurcation
One of the issues is
the bifurcation point of the Maclaurin-like axisymmetric spheroid (which we call
M-spheroid hereafter)
and the Jacobi-like
triaxial ellipsoid (hereafter J-ellipsoid) under the influence of the ring's gravity. 
It is well-known that
the classical Maclaurin spheroid with uniform density and uniform rotation
has the eccentricity $e$ of its meridional section that spans from zero to unity.
An object with $e=0$ corresponds to a non-rotating object, while $e=1$ corresponds to an
infinitely thin disk. For $e\ge 0.81267$, there exists another triaxial ellipsoid with
uniform rotation and with the same angular momentum as Maclaurin spheroid,
which is called Jacobi ellipsoid \citep{2013srlm.book.....L,1969efe..book.....C}. 
It is not known
how this classical picture is modified when the spheroid/ellipsoid is surrounded
by a massive ring. Do we always have a bifurcation point of the axisymmetric
M-spheroid and the triaxial J-ellipsoid for any parameter
choice of the massive ring?
%-- secular stability
Another issue is the secular stability of the M-spheroids. Beyond the
bifurcation point ($e=0.81267$), the classical Maclaurin spheroid has a larger
energy than the Jacobi ellipsoid with the same angular momentum. When viscous
dissipation is taken into account, the Maclaurin spheroid becomes secularly unstable
and evolves towards the Jacobi ellipsoid by preserving the angular momentum \citep{1963ApJ...137..777R,1973ApJ...181..513P}.
When the tidal field of the massive ring surrounding the spheroid/ellipsoid, 
do we expect the same secular instability for the axisymmetric configuration
beyond the corresponding bifurcation point?

Although these issues may appear academic at first sight, they are related to the
structure and the stability of various astrophysical objects. For instance, they are
relevant to the formation and stability of rapidly rotating dwarf planets with rings.
A newly born proto-neutron star or a strange star after the iron-core collapse of a dying
massive star or a massive remnant neutron star of a neutron-star binary merger may
rotate rapidly enough to be secularly unstable to viscous dissipation.  These
may be potential sources of quasi-monochromatic gravitational waves
which are detectable by the network of ground-based gravitational
wave observatories. When they
are surrounded by a massive torus of fallback matter or of merger debris,
how does this picture change?
If the dark matter distribution around some of the elliptical galaxies or around 
the bulges of spiral galaxies may be mimicked by that of the ring (as is
suggested by \citealt{2016ARep...60..526K}), do we expect to see the stellar
systems as spheroids or triaxial ellipsoids? These questions
of completely different scales in the Universe may be related to the
simple models considered here.

%%%%%%%%%%%%%%%
\section{Models}
%%%%%%%%%%%%%%%
In the theory of homogeneous ellipsoidal figures \citep{2013srlm.book.....L,1969efe..book.....C}, the isobaric surfaces
are assumed to form concentric ellipsoids and the surface at which
the pressure vanishes is parametrized with the semi-major axes $a_1, a_2, a_3$,
thus the surface is defined as
\beq
	\frac{x^2}{a_1^2} + \frac{y^2}{a_2^2} + \frac{z^2}{a_3^2} = 1.
\eeq
Here we employ a Cartesian coordinate $(x, y, z)$ whose origin
is at the center of mass and the coordinate axes are along the semi-major axes.
Without loss of generality, we may assume $a_3\le a_2 \le a_1$ for the oblate
cases we are interested in here.
%%------------------------ ring potential
\subsection{Ring potential}
\begin{figure}
	\includegraphics[width=\columnwidth]{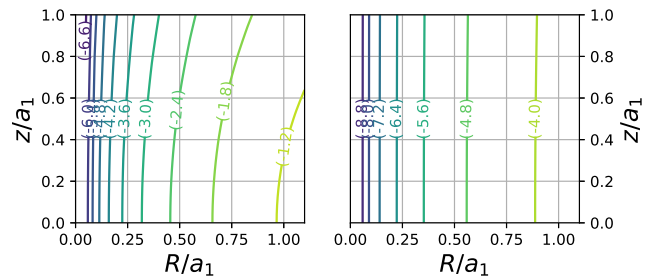}
    \caption{Truncation error of the series expansion of the ring potential
    (Eq.\ref{eq: PhiOfull}) at $n=1$. The contours of the logarithm of the ratio
    $\frac{\sum_{n=2}^{N}\phi_n}{\sum_{n=0}^1\phi_n}$ are shown in the
    $(R, z)$-plane. $\phi_n$ is the n-th term of Eq.(\ref{eq: PhiOfull}).
       The left panel is for $R_o/a_1=2$ and the right one is for $R_o/a_1=10$.
    }
    \label{fig: ring potential}
\end{figure}

The gravitational potential of a thin ring of mass $m_o$
and radius $R_o$, which is a solution of a Poisson's equation, 
is given as a series \citep{2007AmJPh..75..724S},
\beq
	\Phi_o = -Gm_o\sum_{n=0}^\infty \frac{(4n-1)!!}{2^{2n}(n!)^2}
	\frac{(RR_o)^{2n}}{(R^2 + R_o^2 + z^2)^{2n+\frac{1}{2}}},
	\label{eq: PhiOfull}
\eeq
where the cylindrical coordinate $(R,z)$ is used whose $z$-axis coincides with
the rotational axis of the ellipsoid.
To make the solutions semi-analytic, we truncate the ring potential at the
quadratic order in $R$ and $z$, thus $n=0, 1$ in the series. Then the ring potential is
approximated as
\beq
	\Phi_o = -\frac{Gm_o}{R_o}\left[
	1 + \frac{1}{4}\left(\frac{R}{R_o}\right)^2 -\frac{1}{2}\left(\frac{z}{R_o}\right)^2
	\right].
	\label{eq: PhiOquad}
\eeq
The same expression is obtained by \cite{2010Ap.....53..189K} by a-priori
assumption of the quadratic form. 
In Fig.\ref{fig: ring potential}, the truncation error of Eq.(\ref{eq: PhiOfull}) is shown
as contour plots in $(R, z)$-plane. The error is estimated as
\beq
	{\rm error} = \frac{\sum_{n=2}^{N}\phi_n}{\sum_{n=0}^1\phi_n},
	\quad \phi_n \equiv  \frac{(4n-1)!!}{2^{2n}(n!)^2}
	\frac{(RR_o)^{2n}}{(R^2 + R_o^2 + z^2)^{2n+\frac{1}{2}}},
\eeq
where $N$ is a sufficiently large integer. We adopt $N=50$ here.
The coordinate is normalized by the longest semi-major axis $a_1$ and the logarithm of the error is shown here. In an ellipsoid ($0\le R/a_1, z/a_1 \le 1$), the error is largest
at the equatorial surface. For $R_o/a_1=2$, the largest error amounts to a few percent,
while the largest error is less than $10^{-3}$ percent for $R_o/a_1=10$. As is expected,
the truncation gives the better approximation for the larger $R_o$.

%-- error of truncation in ring potential
%

%%-------------------------------------
\subsection{Ellipsoidal equations}
The hydrostatic equation of stationary configuration rotating with
the angular frequency $\Omega$ reads
\beq
	\frac{1}{\rho} \del{x_j}p = -\del{x_j}\Phi +x_j\Omega^2(1-\delta_{j3}) \quad j=1, 2, 3,
\eeq
where $x_1=x, x_2=y, x_3=z$. $\Phi$ is the total gravitational potential.
The first integral form of the equation is
\beq
	\frac{p}{\rho} = -\Phi +\frac{1}{2}(x^2+y^2)\Omega^2 + C,
	\label{eq: 1st integral}
\eeq
where $C$ is a constant.

The internal potential $\Phi_i$ of the self-gravitating uniform density ellipsoid is expressed 
by a simple quadratic form 
\beq
	\Phi_i = -\pi G\rho\left[A - A_1x^2 - A_2y^2 - A_3z^2\right].
	\label{eq: Phii}
\eeq
The coefficients $A_j$ are given by
\beq
	A_j = a_1a_2a_3\int_0^\infty \frac{du}{(a_j^2+u)\Delta},~A=\sum_{j=1}^3a_j^2A_j,
	\label{eq: A-coefficients}
\eeq
where $\Delta^2 = (a_1^2+u)(a_2^2+u)(a_3^2+u)$.

Since the potential of the ring mass is given by Eq.(\ref{eq: PhiOquad}), the total
potential $\Phi = \Phi_i + \Phi_o$ is written as
\begin{eqnarray}
	\Phi &=& -\frac{3GM}{4a_1a_2a_3}\left[
		A + \frac{4a_2a_3m_o}{3\lambda M}\right.\nonumber\\
&&\left.	-\left(A_1-\frac{a_2a_3m_o}{3\lambda^3 a_1^2 M}\right)x^2
		-\left(A_2-\frac{a_2a_3m_o}{3\lambda^3 a_1^2 M}\right)y^2\right.\nonumber\\
&&		\left.
		-\left(A_3+\frac{2a_2a_3m_o}{3\lambda^3 a_1^2 M}\right)z^2
	\right]\nonumber\\
	&\equiv& -\frac{3GM}{4a_1a_2a_3}\left[
	A' - A'_1x^2 -A'_2 y^2 -A'_3 z^2
	\right].
	\label{eq: Phi total}
\end{eqnarray}
Here $\lambda=R_o/a_1$. 

From the condition that the surface of the ellipsoid coincides
with the equipotential surface, we have
\beq
	a_1^2\left(A'_1-\frac{\Omega^2}{2\pi G\rho}\right)
	= a_2^2\left(A'_2-\frac{\Omega^2}{2\pi G\rho}\right)
	= a_3^2 A'_3.
	\label{eq: equality of As 1}
\eeq
Then the equalities,
\beq
	a_1^2a_2^2(A'_1-A'_2) + (a_1^2-a_2)a_3^2A'_3=0
	\label{eq: equality of As 2}
\eeq
and
\beq
	\frac{\Omega^2}{2\pi G\rho} = \frac{a_1^2A'_1-a_2^2A'_2}{a_1^2-a_2^2}
	=  \frac{a_1^2A'_1-a_3^2A'_3}{a_1^2}
	= \frac{a_2^2A'_2-a_3^2A'_2}{a_2^2},
	\label{eq: equality of As 3}
\eeq
hold. From the definition of $A_i$'s and $A'_i$'s above, we may deduce
from Eq.(\ref{eq: equality of As 2}) that
\begin{eqnarray}
&&(a_2^2-a_1^2)\left[
\int_0^\infty\left(\frac{a_1^2a_2^2}{(a_1^2+u)(a_2^2+u)}-\frac{a_3^2}{a_3^2+u}
\right)\frac{du}{\Delta}\right.\nonumber\\ 
&&\left. - \frac{2}{3}\lambda^{-3}\frac{m_o}{M}\left(\frac{a_3}{a_1}\right)^2
\right] = 0
\label{eq: MacJac0}
\end{eqnarray}

This equation results in two equilibrium sequences. One is the case that
$a_1=a_2$ holds. This corresponds to axisymmetric spheroids. We call
them M-spheroid sequence. The other is when the expression in the square
bracket vanishes. It corresponds to sequences of
non-axisymmetric triaxial ellipsoids. We call them J-ellipsoids.
The functional relation of $\Omega$ and the normalized principal axes
for an M-spheroid sequence is given by the second or third equality
of Eq.(\ref{eq: equality of As 3}). For a J-ellipsoid sequence, the  
expression in the square bracket of Eq.(\ref{eq: MacJac0}) is set zero
and solved for $a_3$ when $a_1$ and $a_2$ is specified. The values of $a_j$'s
are substituted to Eq.(\ref{eq: equality of As 3}) to obtain $\Omega$.
The bifurcation point of M-spheroid and J-ellipsoid sequences
corresponds to the special solution of this equation for J-ellipsoid
by setting $a_1=a_2$.

Following the conventional treatment of the ellipsoidal figures, we parametrize
the sequences by the eccentricity $e$ of the meridional section of the ellipsoids,
\beq
	e = \sqrt{1-\left(\frac{a_3}{a_1}\right)^2}.
\eeq

%Hereafter we normalize the length scale with $a_1$. We define
%$u=a_1\xi$, $\alpha_2 = a_2/a_1$, $\alpha_3=a_3/a_1$

\subsection{Expressions of energy and Angular momentum}
The rotational kinetic energy $T$ of a uniformly rotating ellipsoid is obtained by
\beq
	T = \frac{1}{2}I_z\Omega^2,
\eeq
where $I_z$ is the moment of inertia and is computed by
\beq
	I_z = \frac{4}{15}\pi\rho a_1a_2a_3(a_1^2+a_2^2).
\eeq
The angular momentum $J$ is simply given by $J=I_z\Omega$.

The gravitational energy $W$ is given by,
\beq
	W = W_i + W_o = \frac{1}{2}\int\rho\Phi_i dV + \int\rho\Phi_o dV.
\eeq
Notice that the potential $\Phi_i$ is produced by the mass
of the ellipsoid itself, thus the contribution from the self-gravity
is multiplied by the $1/2$ factor to cancel the double count, 
while $\Phi_o$ is the external
potential created by the ring, which does not require the factor.
The self-gravity term $W_i$ is computed as,
\beq
	W_i = -\frac{8}{15}\pi^2 G\rho a_1 a_2 a_3 A,
\eeq
while the ring contribution is,
\beq
	W_o = -\frac{Gm_o}{R_o}\frac{4\pi\rho}{3}a_1a_2a_3\left[
	1 + \frac{1}{20}\frac{a_1^2+a_2^2}{R_o^2} - \frac{1}{10}\frac{a_3^2}{R_o^2}.
	\right]
	\label{eq: Wo}
\eeq
The total energy of an ellipsoid is $E=T+W$.

\subsection{Numerical treatment}
To obtain the functional relation of $\Omega$ and the principal axes
of the ellipsoid, we need to evaluate elliptic integrals appearing in $A_j$'s.
More specifically, the integral we use has the form,
\beq
	R_J(x,y,z,p) = \frac{3}{2}\int_0^\infty\frac{du}{(u+p)\sqrt{(u+x)(u+y)(u+z)}}.
\eeq
This is one of Carlson's elliptic integrals \citep{Carlson1979, Carlson-Notis1981}.
We use a {\tt python} version of TOMS577 library to compute this integral.
\footnote{\url{https://people.math.sc.edu/Burkardt/py_src/toms577/toms577.html}}

%%%%%%%%%%%%%%%%%%%%%%%%
\section{Results}
%%%%%%%%%%%%%%%%%%%%%%%%
\subsection{Angular frequency}

As is seen in Eq.(\ref{eq: Phi total}),(\ref{eq: equality of As 3}),(\ref{eq: MacJac0}),
the sequences of $\Omega^2$ as a function of the eccentricity $e$
are characterized by the dimensionless parameter $\nu\equiv \lambda^{-3} m_o/M$.
From the expression of the parameter,
\beq
	\nu = \lambda^{-3}\frac{m_o}{M} =\left(\frac{R_o}{a_1}\right)^{-3}\frac{m_o}{M}
	 = \frac{\frac{Gm_o}{R_o^3}a_1^2}{\frac{GM}{a_1}} \sim \frac{\Phi_o(a_1)}{\Phi_i(a_1)},
\eeq
its physical significance is the relative strength of the ring's gravitational field compared
with that of the ellipsoid at the surface.

\begin{figure}
	\includegraphics[width=\columnwidth]{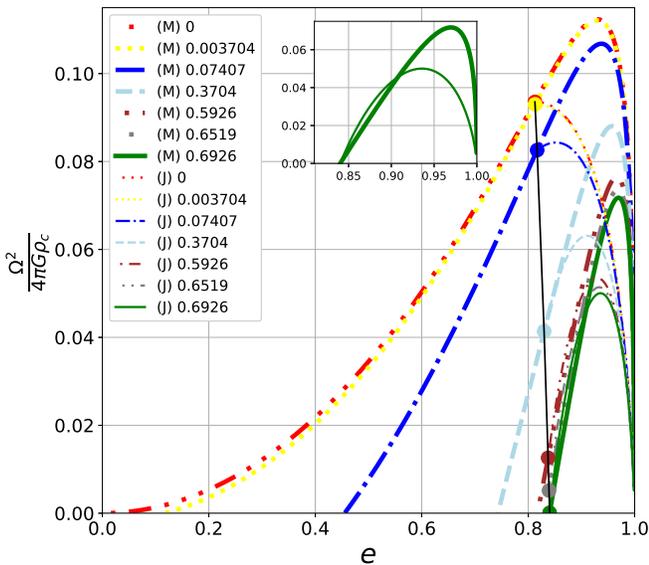}
    \caption{The angular frequency squared $\Omega^2$, normalized by
    $4\pi G\rho_c$, is plotted as a function of the eccentricity $e=\sqrt{1-\left(a_3/a_1\right)^2}$. 
    The averaged axis $\bar{a}$ is defined as $\bar{a}=(a_1a_2a_3)^{1/3}$.
    %The inset shows the enlarged picture of the curves around the bifurcation points.
    The thick curves with a label (M) are M-spheroids, while those with (J) are J-ellipsoids.
    The number in the legend is the value of $\nu$, where $\nu=0$ curves are the original
    Maclaurin and the Jacobi sequences.
    The filled circles and the solid curve connecting them mark the bifurcation points
    of M-spheroids and J-ellipsoids.
    The inset shows the M-spheroids and the J-ellipsoids sequences of $\nu=0.6926$,
    for which $\Omega=0$ at the bifurcation.
    }
    \label{fig: Omega2}
\end{figure}

In Fig.\ref{fig: Omega2}, sequences of $\Omega^2$ for the M-spheroids and the J-ellipsoids are plotted as functions of eccentricity. For a given line style, the thicker curve
corresponds to the M-spheroid sequence. The bifurcation points for each $\nu$ are marked
with the circles. 
%-- Omega^2
We see that the normalized frequency decreases as the parameter $\nu$ increases.
Also, the M-spheroid sequence in general does not have a $e\to 0$ limit.
This is in accord with the result by \cite{2016ARep...60..526K}, and
in contrast to the Maclaurin spheroid embedded in a massive spherical
halo \citep{1984A&A...133..369R}, for which the sequences always seem to extend
to $e\to 0$.

For a small value of $\nu$ ($\le 0.3108$), the Jacobi-like model always 
has a lower $\Omega$ than
the M-spheroid model with the same $e$. As $\nu$ increases, however, there appears
a range of $e$ within which $\Omega$ of J-ellipsoids becomes larger
than that of M-spheroids. This is clearly seen in the inset.

As $\nu$ increases, the normalized $\Omega^2$ decreases, though
the eccentricity at the bifurcation changes little. Overall, the eccentricity 
of the bifurcation point shifts at most 3.4\%. This is compared with the
result by \cite{1967PASJ...19..242M} who showed
the bifurcation point of the Maclaurin spheroid embedded in a concentric
spheroidal halo shifts more than 10\% as the density of the halo doubles the
central spheroid.

Another remarkable trait of the bifurcation is that for $\nu>0.6926$, the bifurcation point
vanishes at which an M-spheroid and a J-ellipsoid coincide.
%-- Omega^2
\begin{figure}
	\includegraphics[width=\columnwidth]{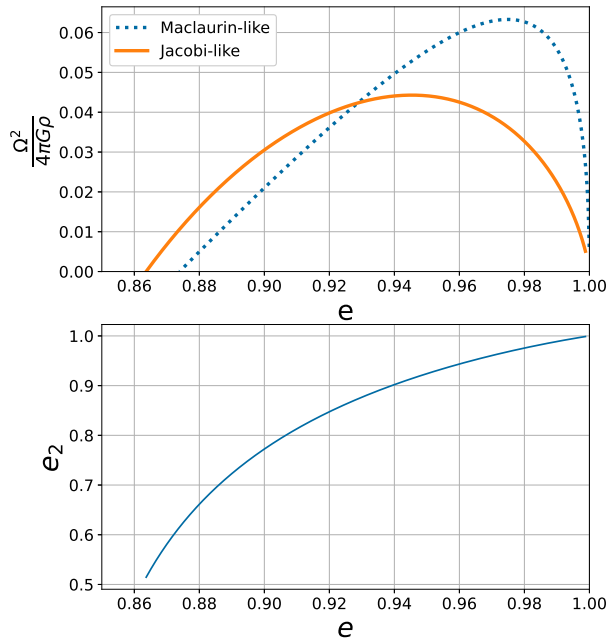}
    \caption{Top: An M-spheroid and a J-ellipsoid sequence for $\nu>0.6926$.
    Here $m_0=3$ and $\lambda=1.5$, thus $\nu=0.8889$.
    Bottom: The eccentricity $e$ in the x-z-section versus that of the x-y-section
    $e_2$ for the J-ellipsoid sequence above.
    }
    \label{fig: large-nu}
\end{figure}
For this case, shown in Fig.\ref{fig: large-nu}, a J-ellipsoid appears at a lower eccentricity than an M-spheroid, and the former can be a configuration with zero rotational frequency.
The intersection of the two curves is not a bifurcation point of the M-spheroid
and the J-ellipsoid sequences. This is seen in the lower panel of Fig.\ref{fig: large-nu},
where the eccentricities of the J-ellipsoids in x-z- and x-y-sections
are compared. The eccentricity in the x-y-section, $e_2$, is defined as $e_2=\sqrt{1-(a_2/a_1)^2}$. For a M-spheroid, $e_2=0$ by definition. Therefore the
two sequences do not share the $e_2$ value at the intersection point of the
two curves, though
the two sequences of ellipsoids would have shared it at a genuine bifurcation point.

%---------------------------------------------------------
\subsection{Secular stability of M-spheroid}

\begin{figure}
	\includegraphics[width=\columnwidth]{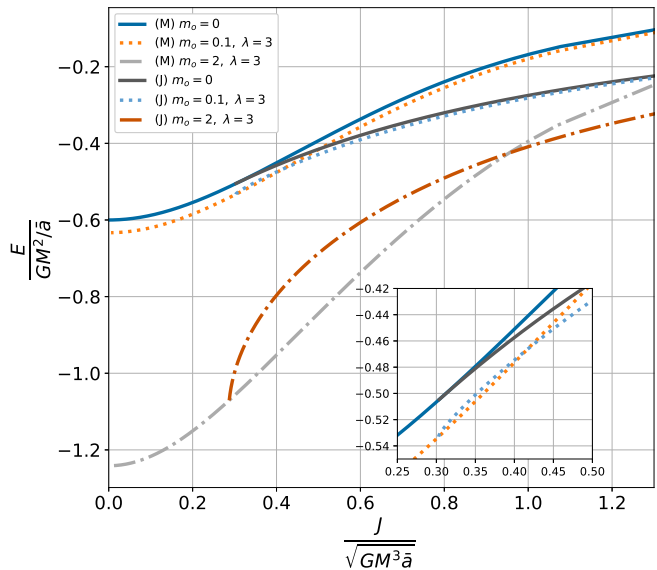}
    \caption{The energy as a function of the angular momentum of the ellipsoids.
    The energy $E$ is normalized by $GM^2/\bar{a}$ and the angular momentum
    is normalized by $\sqrt{GM^3\bar{a}}$. The inset shows the enlarged picture
    around the bifurcation points for $m_o=0~{\rm and}~0.1$ models.
    }
    \label{fig: Energy}
\end{figure}

Notice that the parameters $\lambda$ and $m_o/M$ are not degenerate
in the expression of the energy (see Eq.(\ref{eq: Wo})), thus the models
are not represented by the single parameter $\nu$, but we need to specify
both $m_o/M$ and $\lambda$.
In Fig.\ref{fig: Energy}, we plot the energy as a function of the angular
momentum for the M-spheroid and the J-ellipsoid sequences. The parameters
are chosen so that $\nu<0.6926$. The enlarged view of the sequences around
the bifurcation points are shown in the inset. The original Maclaurin spheroid
always has a higher energy than the Jacobi ellipsoid with the same angular
momentum. Therefore the Maclaurin spheroids are secularly unstable beyond
the bifurcation point. Viscous energy dissipation would drive the Maclaurin
spheroid to the lower-energy Jacobi ellipsoid with the same angular momentum.
When the gravitational field of the ring is turned on, the J-ellipsoid may
have higher energy than the M-spheroid around the bifurcation
point. The energy of the former becomes lower than the latter for sufficiently
large angular momentum. Thus the M-spheroid is secularly stable
at least around the bifurcation point and the neutral stability point is located
at a higher angular momentum. 
%--E vs J
For the case with $\nu>0.6926$, the J-ellipsoid sequence always has
larger energy than the M-spheroid sequence. Thus the tidal field of the
massive ring stabilizes the M-spheroids against the bar-shaped 
bifurcation to triaxial ellipsoids.

%--critical eccentricity of M-spheroid for secular instability
\begin{figure}
	\includegraphics[width=\columnwidth]{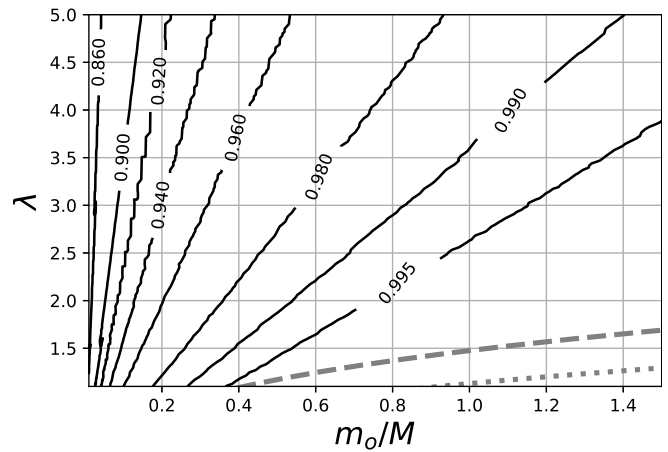}
    \caption{Contour map of the eccentricity of M-spheroids
    at their neutral stability. The dashed line is the curve with
    $\nu=0.3108$, while the dotted line is for $\nu=0.6926$.
       }
    \label{fig: ecritical}
\end{figure}
In Fig.\ref{fig: ecritical}, the eccentricity $e$ of the M-spheroid 
at which its total energy equals that of the J-ellipsoid with
the same angular momentum (it should be noted that the ellipsoid 
does not necessarily share the eccentricity
with the M-spheroid). Below a contour curve, the M-spheroid with
the eccentricity lower than the labeled value is secularly
unstable to viscous dissipation, since its energy is higher than the J-ellipsoid.
It should be noted that the classical Maclaurin spheroid becomes secularly
unstable for $e\ge 0.81267$. The gravity of the ring makes the M-spheroid 
more stable and the effect is enhanced as $\nu=\lambda^{-3}m_o/M$ becomes
larger (towards the bottom right corner of the figure).

%----
\section{Summary and discussion}
We studied the uniformly rotating homogeneous ellipsoidal
sequences under the influence of the gravitational field
of a massive concentric ring. The ring potential is truncated
at the quadrupole term in order for the study to be
semi-analytic. The resulting expressions for the various
physical quantities have correction terms determined by
the mass and the radius of the ring. When the mass is
reduced to zero or the radius is set to be infinite, we
recover the classical Maclaurin spheroids and the Jacobi
ellipsoids. The existence of the massive ring modifies 
the bifurcation and the secular stability of the ellipsoids. 
The meridional section of an M-spheroid with zero angular frequency
has a finite eccentricity $e$ due to the gravitational attraction
of the ring mass. Thus the sequence is not extended to 
$e=0$ limit. If the parameter $\nu$ which measures the gravitational
field of the ring is smaller than $\nu=0.6926$, a J-ellipsoid bifurcates
from an M-spheroid at some value of $e$ as is seen in the classical
Maclaurin sequence. Beyond this threshold, the bifurcation point
vanishes, and the J-ellipsoid exists at a lower $e$ than the M-spheroid
appears. Examining the total energy of the M-spheroid and
the J-ellipsoid with the same angular momentum, the M-spheroid
is not secularly unstable to viscous dissipation around the bifurcation
point, in contrast to the classical Maclaurin spheroid which is destabilized
beyond the bifurcation point of Jacobi ellipsoids. When parametrized
by the eccentricity, the stabilization occurs
before the rotational frequency of the J-ellipsoid may exceed that of
the M-ellipsoid ($\nu\le 0.3108$).

It may seem counter-intuitive that a non-rotating object with an
external axisymmetric ring source develops
a triaxial shape as is seen in Fig.\ref{fig: Energy}. This occurs when the gravity
of the ring is comparable to the self-gravity of the ellipsoid ($\nu >0.6926$).
On the equatorial plane ($z=0$), the ring's potential (Eq.\ref{eq: PhiOfull})
is quadratic in the cylindrical radial coordinate. Therefore it may be effectively
identified as the rotational term in Eq.(\ref{eq: 1st integral}). In other words, 
the gravitational attractive force of the circum-object ring induces an effective centrifugal effect
on the ellipsoid. It results in the appearance of the Jacobi-like triaxial ellipsoid
even if the object's rotational frequency vanishes.

%%%%%%%%%%%%%%%%%
%\section*{Acknowledgements}
%%%%%%%%%%%%%%%%%

%%%%%%%%%%%%%%%%%
%\section*{Data Availability}
%%%%%%%%%%%%%%%%%%
%No new data were generated or analysed in support of this research.
%%%%%%%%%%%%%%%%%%%%% REFERENCES %%%%%%%%%%%%%%%%%%


\bibliographystyle{apsrev}
\bibliography{ellip} % if your bibtex file is 

\end{document}